\documentclass[]{aastex631}

\shorttitle{GRB\,180128A}
\shortauthors{Trigg et al.}

\graphicspath{{./}{figures/}}

          \font\sixrm=cmr6       
                     
\begin{document}

\newcommand{\fermi}{\textit{Fermi}-GBM}
\newcommand{\fluxcgs}{ergs~s$^{-1}$~cm$^{-2}$}
\newcommand{\lumcgs}{ergs~s$^{-1}$}
\newcommand{\fluengs}{ergs~cm$^{-2}$}
\newcommand{\gev}{GeV~}
\title{GRB\,180128A: A Second Magnetar Giant Flare Candidate from the Sculptor Galaxy}

\newcommand{\zw}[1]{\textcolor{cyan}{#1}}

\author[0009-0006-8598-728X]{Aaron C. Trigg}
\affiliation{Department of Physics \& Astronomy, Louisiana State University, Baton Rouge, LA 70803, USA}
\author[0000-0002-2942-3379]{Eric Burns}
\affiliation{Department of Physics \& Astronomy, Louisiana State University, Baton Rouge, LA 70803, USA}
\author[0000-0002-7150-9061]{Oliver J.~Roberts}
\affiliation{Science and Technology Institute, Universities Space and Research Association, 320 Sparkman Drive, Huntsville, AL 35805, USA.}
\author[0000-0002-6548-5622]{Michela Negro}
\affiliation{Department of Physics \& Astronomy, Louisiana State University, Baton Rouge, LA 70803, USA}
\author[0000-0002-2208-2196]{Dmitry S. Svinkin}
\affiliation{Ioffe Institute, 26 Politekhnicheskaya, St. Petersburg, 194021, Russia}
\author[0000-0003-4433-1365]{Matthew G. Baring}
\affiliation{Department of Physics and Astronomy - MS 108, Rice University, 6100 Main Street, Houston, TX 77251-1892, USA}
\author[0000-0002-9249-0515]{Zorawar Wadiasingh}
\affiliation{Astrophysics Science Division, NASA/GSFC, Greenbelt, MD 20771, USA}
\affiliation{Department of Astronomy, University of Maryland, College Park, MD 20742, USA}
\affiliation{Center for Research and Exploration in Space Science and Technology, NASA/GSFC, Greenbelt, MD 20771, USA}
\author[0000-0002-6870-4202]{Nelson L. Christensen}
\affiliation{Universit\'e C\^ote d’Azur, Observatoire de la C\^ote d’Azur, CNRS, Artemis, 06304 Nice, France}
\author[0000-0002-8977-1498]{Igor Andreoni}
\altaffiliation{Neil Gehrels Fellow}
\affil{Joint Space-Science Institute, University of Maryland, College Park, MD 20742, USA}
\affil{Department of Astronomy, University of Maryland, College Park, MD 20742, USA}
\affil{Astrophysics Science Division, NASA Goddard Space Flight Center, Mail Code 661, Greenbelt, MD 20771, USA}
\author[0000-0003-2105-7711]{Michael S. Briggs}
\affiliation{Department of Space Science, University of Alabama in Huntsville, Huntsville, AL 35899, USA}
\author[0000-0002-7574-1298]{Niccol\`o Di Lalla}
\affiliation{Department of Physics and Kavli Institute for Particle Astrophysics and Cosmology, Stanford University, Stanford, CA 94305, USA}
\author[0000-0002-2208-2196]{Dmitry D. Frederiks}
\affiliation{Ioffe Institute, 26 Politekhnicheskaya, St. Petersburg, 194021, Russia}
\author[0000-0003-3668-1314]{Vladimir M. Lipunov}
\affiliation{Department of Physics, Lomonosov Moscow State University, Sternberg Astronomical Institute,
119991, 13, Univeristetskij Prospekt, Moscow, Russia}
\author[0000-0002-5448-7577]{Nicola Omodei}
\affiliation{Department of Physics and Kavli Institute for Particle Astrophysics and Cosmology, Stanford University, Stanford, CA 94305, USA}
\author[0000-0001-9477-5437]{Anna V. Ridnaia}
\affiliation{Ioffe Institute, 26 Politekhnicheskaya, St. Petersburg, 194021, Russia}
\author[0000-0002-2149-9846]{Peter Veres}
\affiliation{Department of Space Science, University of Alabama in Huntsville, Huntsville, AL 35899, USA}
\author[0000-0002-3942-8341]{Alexandra L. Lysenko}
\affiliation{Ioffe Institute, 26 Politekhnicheskaya, St. Petersburg, 194021, Russia}

\begin{abstract}

Magnetars are slowly rotating neutron stars that possess the strongest magnetic fields ($10^{14}-10^{15}~\mathrm{G}$) known in the cosmos. They display a range of transient high-energy electromagnetic activity. The brightest and most energetic of these events are the gamma-ray bursts (GRBs) 
 known as magnetar giant flares (MGFs), with isotropic energy $E\approx10^{44}-10^{46}~\mathrm{erg}$. There are only seven detections identified as MGFs to date: three unambiguous events occurred in our Galaxy and the Magellanic Clouds, and the other four MGF candidates are associated with nearby star-forming galaxies. As all seven identified MGFs are bright at Earth, additional weaker events remain unidentified in archival data. We conducted a search of the \textit{Fermi}~Gamma-ray Burst Monitor (GBM) database for candidate extragalactic MGFs and, when possible, collected localization data from the Interplanetary Network (IPN) satellites. Our search yielded one convincing event, GRB\,180128A. IPN localizes this burst with NGC\,253, commonly known as the Sculptor Galaxy. This event is the second MGF in modern astronomy to be associated with this galaxy and the first time two bursts are associated with a single galaxy outside our own. Here, we detail the archival search criteria that uncovered this event and its spectral and temporal properties, which are consistent with expectations for a MGF. We also discuss the theoretical implications and finer burst structures resolved from various binning methods. Our analysis provides observational evidence for an eighth identified MGF.
\end{abstract}

\keywords{Gamma-ray bursts(629) --- Magnetars(992)}

\section{Introduction} \label{sec:intro}

A magnetar is a type of neutron star (NS) characterized by an extremely strong magnetic field, $>10^{14}~\mathrm{G}$ \citep{1984Ap&SS.107..191U,duncan1992formation,Paczynski:1992zz,1995MNRAS.275..255T,1996ApJ...473..322T,2003ApJ...588L..93K}. There are about thirty identified magnetars within our Galaxy \citep{Olausen_2014} and are associated with star-forming regions \citep{Gaensler2004AdSpR..33..645G}. These objects display a wide variety of high-energy transient activity: from short (sub-second) GRBs to burst forests with hundreds or thousands of bursts within tens of minutes, to MGFs, the most energetic class of transient events from magnetars \citep[see, e.g.,][for a recent review]{kaspi2017}.

MGFs are characterized by an intense initial pulse (``spike'') typically showing a millisecond-long rise time, peak energy in the gamma-ray band ($\sim$MeV), and total isotropic-equivalent energy $E_{\mathrm{iso}}\gtrsim10^{44}~\mathrm{erg}$. In the three most proximate MGFs, observations show that a long, decaying tail with a duration of several hundred seconds follows this pulse, the intrinsic energy of which is of the order of a few $10^{44}~\mathrm{erg}$. The rotation of the NS periodically modulates this tail. Given the extremely high peak luminosities of the initial spikes, detection of these emissions is possible for magnetars located in galaxies of the local group up to a distance of $\sim10~\mathrm{Mpc}$ \citep{burns2021identification} by sensitive instruments such as \fermi. Furthermore, it has been shown in previous studies that a fraction ($\sim2\%$) of MGFs from nearby galaxies masquerade as short GRBs \citep{palmer2005giant,ofek2007soft,hurley2011short,Svinkin2015MNRAS.447.1028S,burns2021identification}. In contrast, at extragalactic distances, the modulated tail indicative of a MGF is too faint to be observed with current instruments due to limitations in their sensitivity \citep{Hurley_2005Natur.434.1098H}. Given these current limitations, the best method to identify extragalactic MGF candidates is by looking for events displaying the distinct property of the prompt gamma-ray emission that are also spatially aligned with nearby star-forming galaxies.
 
The current sample of MGFs counts seven events, three of which happened locally (in the Milky Way and the Large Magellanic Cloud; \citealt{mazets1979observations,hurley1999giant,mazets1999AstL...25..635M,palmer2005giant,2007AstL...33...19F}) and, due to their exceptional brightness, saturated all observing instruments at that time.
The remaining four events, GRB\,051103, GRB\,070201, GRB\,070222, and GRB\,200415A, were found to have spatial alignment in 2D with nearby star-forming galaxies M81, M31, M83, and NGC 253 (also called Sculptor galaxy), respectively \citep{ofek2006short,2007AstL...33...19F,mazets2008giant,2008ApJ...681.1464O,hurley2010new,svinkin2021bright,2021Natur.589..207R,burns2021identification}. Despite this small sample of events, \cite{burns2021identification} recently reported a very high intrinsic volumetric rate of $R_{\hbox{\sixrm MGF}} =3.8^{+4.0}_{-3.1}\times10^{5}~\mathrm{Gpc^{-3}~yr^{-1}}$, supporting the idea that a commonly occurring progenitor, such as regular core-collapse supernovae (SN), is behind the origin of magnetars and that some may produce multiple MGFs, which would help inform our understanding of the mechanisms that cause them.

Section \ref{sec:id} relates our method in identifying and localizing this detection. A detailed analysis of \fermi\,data for the newly found event, and a side-by-side comparison with the other MGF associated with NGC~253, GRB\,200415A, and a likely NS merger GRB\,150101B follows in Sections \ref{sec:gbmtempanal}. Section \ref{sec:LATana} covers the \textit{Fermi} Large Area Telescope (LAT) data analysis. Sections \ref{sec:Optical}, \ref{sec:GW}, and \ref{sec:radio} outline our search for detections of this event in other signals. In Section \ref{sec:GBMDisInc}, we discuss the relativistic wind model for MGFs and how it relates to GRB\,180128A, as well as the multi-pulse characteristics seen in the lightcurves of this new MGF candidate and three others, and the implications of finding a second burst localized to NGC\,253. Finally, we conclude our findings in Section \ref{sec:conclude}.

\section{MGF Identification} \label{sec:id}

The \textit{Fermi}~GBM consists of 12 uncollimated Thallium-doped Sodium Iodide (NaI) detectors and two Bismuth Germanate (BGO) detectors. The NaI and BGO detectors arranged opposite sides of \textit{Fermi} (n0-n5 and b0 on one side and n6-nb and b1 on the other), with the NaI detectors oriented to observe the entire unocculted sky. The effective spectral ranges of the NaI and BGO detectors are $\sim$8--900~keV and $\sim$0.25--40~MeV, respectively, resulting in a combined spectral range of $\sim$8~keV to 40~MeV. To date, several catalogs detailing GRB data collected by the \fermi\, \citep{gruber2014fermi,von2014second, bhat2016third, von2020fourth} have been released. The events listed in these catalogs are generally longer and have longer variability timescales than MGFs \citep{gruber2014fermi,von2014second,bhat2016third,von2020fourth}. The key parameters to separate MGFs from other prompt observations are the rise time and peak interval duration, but neither are available in the GRB catalogs. More information about the GBM instrument is available in \citet{meegan2009fermi}.

Extragalactic candidate MGFs can be distinguished from regular short GRBs as the former show a shorter rise time and pulse duration, higher spectral energy peak, and are associated with host galaxies at distances appropriate to produce a given peak luminosity ($L_{\rm{p}}\sim10^{41}-10^{47}$) at Earth. We scour the archival \fermi\,data, as we expect more MGFs ($\sim$2--3) may be present, searching for MGF-like detections characterized by sharp, ms-long rise times and a peak interval duration consistent with the known MGF sample. When data are available, we collect localization data from IPN for the flagged MGF candidates. Here, we describe the data and the methods used to discern MGF candidates from the pool of GBM triggers.

In this work, we utilize the time-tagged events (TTE) data collected in the \textit{Fermi}~GBM catalog until August of 2022. TTE data are generated for each detector with a $2\mu$s temporal resolution, with each photon tagged by the arrival time and one of the 128 energy channels, with separate channels for the NaI and BGO detectors. To analyze the data, we use the GBM~Data~Tools~\citep{GbmDataTools}. We then apply a Bayesian Blocks analysis (BB, \citealt{scargle2013bayesian}) to search for significant pulses in the lightcurves. Since the majority of extragalactic MGFs generally have durations of $<100~\mathrm{ms}$ \citep[][and references therein]{burns2021identification}, our first event selection criterion is to select only the events in the catalog with a $T_{50}$ smaller than 100 ms. The $T_{50}$ is the duration over which 50\% of the burst fluence accumulates. The start of the interval is when 25\% of the total fluence is detected, measured between 50\,keV and 300\,keV. Applying the threshold to $T_{50}$ instead of $T_{90}$ (i.e., the duration during which 5\% to 95\% of the burst fluence accumulates over the same energy range) should prevent the removal of any real, fainter MGFs. We then collect \textit{bcat detector mask}\footnote{\url{https://heasarc.gsfc.nasa.gov/w3browse/fermi/fermigbrst.html}} information to determine the best viewing subset of detectors for a given burst. This information indicates which NaI detectors contribute to ``bcat'' files, which provide basic information on a given burst, such as duration, peak flux, and fluence. We can select the appropriate BGO detector data based on which side (defined above) of \textit{Fermi} has more NaI detectors used in the bcat files. This initial selection yields a list of 137 possible MGF candidates.

Using the default parameters, we apply the BB algorithm in {\tt astropy}\footnote{\url{https://docs.astropy.org/en/stable/api/astropy.stats.bayesian_blocks.html}} to the detector data and select events for which the most significant BB bin is shorter than 20 ms. This value is short enough to discard any longer types of GRBs but still long enough to include the entirety of the initial prompt emission of previously identified MGFs. To ensure consideration of only events with very sharp rise times, we require that the first significant BB bin happens within 10 ms from the most significant BB bin. We define significant bins as those with a signal-to-noise ratio greater than 3.5. After this step, 20 event candidates remain. 
We further remove events with known redshift, as the isotropic luminosity for events at these distances would exclude a MGF origin. We also eliminate the MGF previously identified by \cite{burns2021identification}, leading to a considered sample of 13 MGF candidates.

\begin{figure}[h]
    \centering
    \begin{minipage}[t]{0.47\textwidth}
        \centering
        \includegraphics[width=\linewidth]{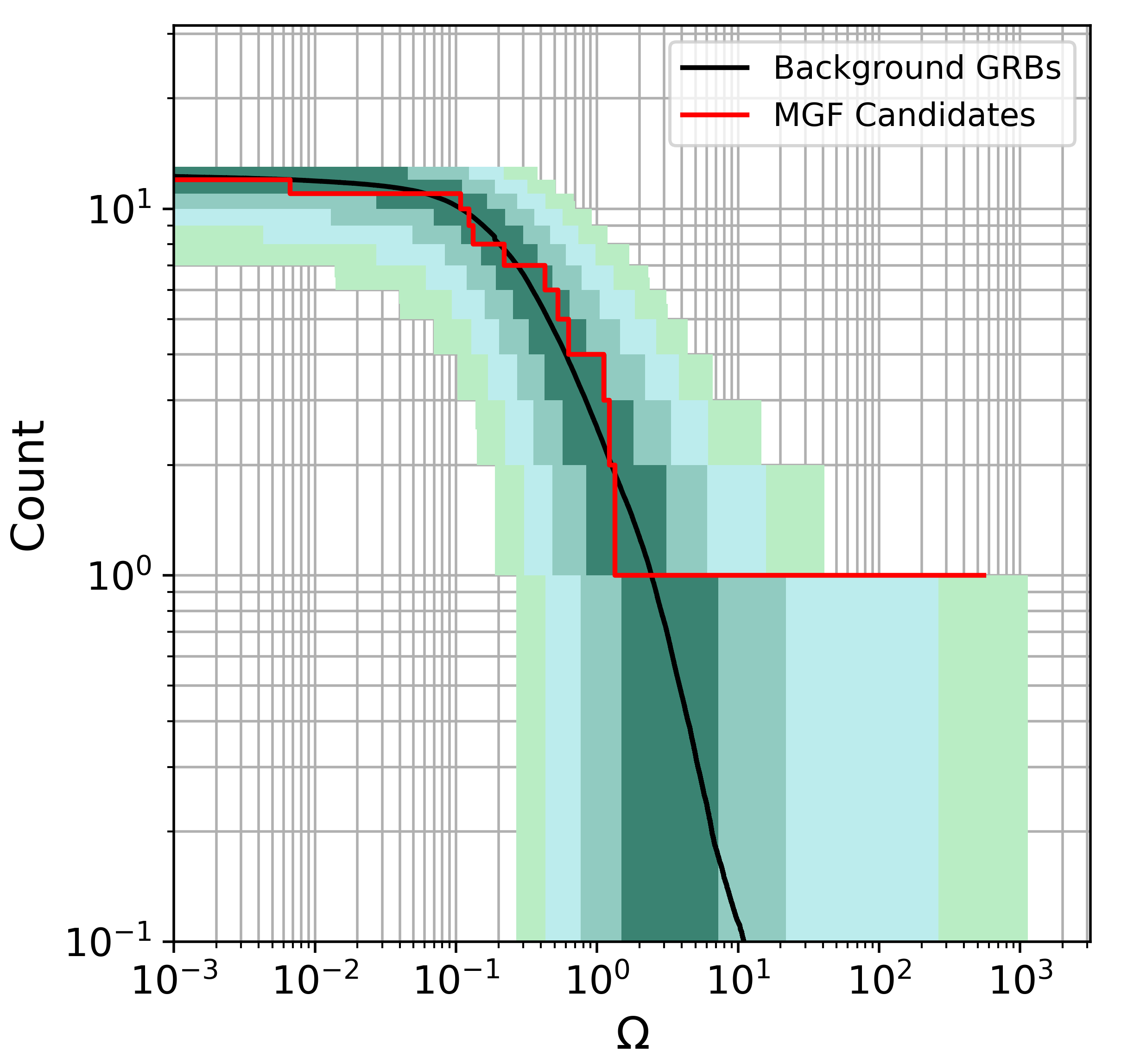}
        \caption{The significance of the selected sample of 13 MGF candidates. $\Omega$ is a ranking statistic representing the believability that a given burst is a giant flare based on 3D spatial agreement with cataloged galaxies, as described in \cite{burns2021identification}. The red line indicates the 13 MGF candidates while the black line represents the background distribution of GRBs. The green-shaded regions represent the 1, 2, 3, and 4$\sigma$ two-sided confidence intervals. There is a single event outlier at 3.3$\sigma$ significance; GRB\,180128A.}
        \label{fig:significance}
    
    \end{minipage}
    \hfill
    \begin{minipage}[t]{0.46\textwidth}
        \centering
        \includegraphics[width=\linewidth]{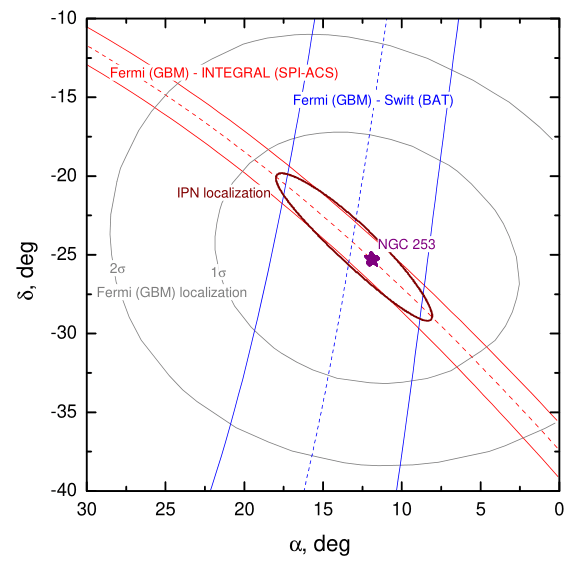}
        \caption{Final IPN localization of GRB\,180128A. The localization defined by the \textit{Fermi-INTEGRAL} (red) and \textit{Fermi-Swift} (blue) annuli. The shown annuli widths have 3$\sigma$ confidence. The overlap of the annuli gives an ellipse (purple) with a 90\% confidence area of 9.3 deg$^2$. The star marks the location of NGC~253. The initial \fermi\, localization and confidence intervals are seen in gray.}
        \label{fig:IPN}
    \end{minipage}
\end{figure}

The autonomous localization capability of \fermi\, is insufficient to robustly associate a GRB to a host galaxy. Thus, IPN constructs annuli for these 13 events. Two of these events were detected only by \fermi\,. Five others have detections by \fermi\, and INTEGRAL \citep{2003SPIE.4851.1336V}, which only constrains the annuli that reduce the GBM localization.

The remaining six events have detections by \fermi\,, INTEGRAL, and \textit{Swift}-BAT \citep{2005SSRv..120..143B}. We note that none of the MGF candidates have \textit{Swift}-BAT localizations. In quantifying the likelihood of these 13 candidates being MGFs from known galaxies, we follow the procedure in \citet{burns2021identification}, which compares two probability density functions (PDFs). The first one, ($P^{GRB}$) represents a distribution of well-localized short GRBs and compares GRB position posteriors against galaxy positions, where galaxies are weighted by distance and their integral star formation rate. Our comparison is run against a background distribution generated by rotating the positions of galaxies from the z0MG catalog \citep{2019ApJS..244...24L} and Atlas of the Local Volume Galaxies \citep[LVG,][]{2013AJ....145..101K}, which then generates the confidence intervals as shown in Figure\,\ref{fig:significance}. The second PDF, $P^{MGF}$, represents the likelihood that a MGF of a given fluence at Earth could originate from a given position. The ranking statistic, $\Omega=4\pi\sum_{i}P_{i}^{GRB}P_{i}^{MGF}/A_{i}$, the product of the probabilities of the PDFs in the \textit{i}$^{th}$ region of the sky of area $A_{i}$, represents the believability that a given burst is a giant flare based on 3D spatial agreement with the cataloged galaxies.

Among all the candidates, the GBM trigger bn180128215, hereafter GRB\,180128A, was immediately identified as a strong candidate, standing out against the background distribution at 3.3$\sigma$ significance. The remainder of this article focuses on the detailed analysis of this event; the population study will be the subject of future work.

Initially, the event was localized to an RA = 12.3 and dec. = -26.1 degrees (J2000)  by \fermi\,, with an average error ellipse radius of 5.7 degrees \citep{connaughton2015localization}, detections of GRB\,180128A also exist for  \textit{Swift}-BAT \citep{2005SSRv..120..143B} outside of its coded field of view, and INTEGRAL SPI-ACS \citep{2003SPIE.4851.1336V}. The IPN refined the localization to 9.3\,deg$^2$, which is centered on NGC\,253, as shown in Figure\,\ref{fig:IPN}. In further investigating this event as having a MGF origin, we perform spectral and temporal analyses for this burst.

\section{Prompt MGF Analysis}
\label{sec:GBManalaysis}

For  GRB\,180128A, we analyze the \fermi\,data and search for signals in the \textit{Fermi} LAT. We then compare the results with those from two other GRBs with a confident progenitor classification, allowing us to compare the characteristics of GRB\,180128A with those of a known MGF and a GRB produced by another source.

\begin{figure}[ht]
    \centering
    \includegraphics[width=.99\textwidth]{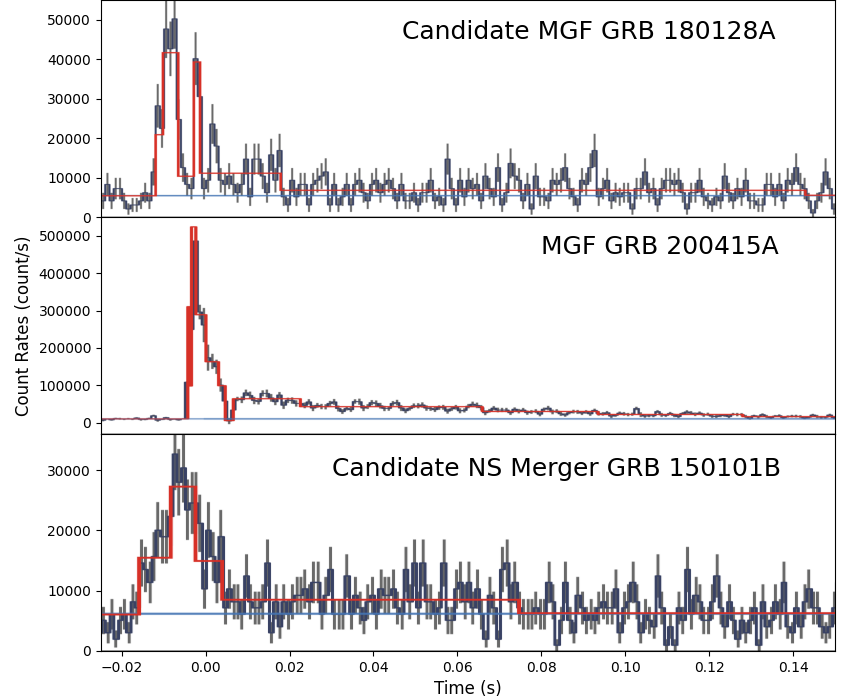}
    \caption{The lightcurves of the three events GRB\,180128A, GRB\,200415A, GRB\,150101B, binned to a temporal resolution of 1~ms for energies from 10--500 keV. The black lines represent the raw data. The grey line shows the energy-integrated background. The red lines show the significance of the pulses above the background (grey line) using a Bayesian Blocks algorithm. At this temporal resolution GRB\,180128A displays two distinct ms peaks.}
    \label{fig:GBM_LC}
\end{figure}

\begin{figure}[ht]
    \centering
    \begin{minipage}[t]{0.95\textwidth}
        \centering
        \includegraphics[width=\linewidth]{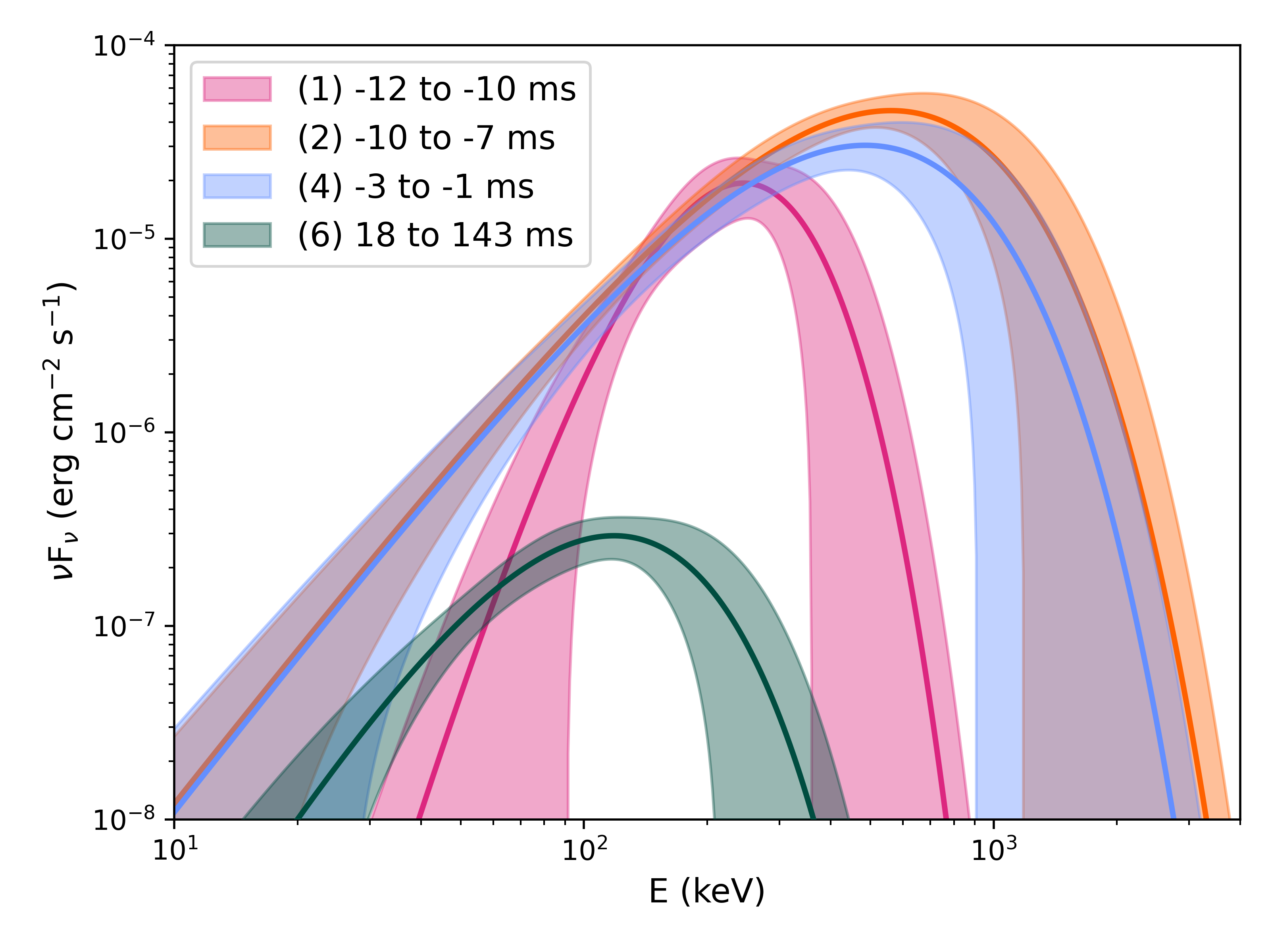}
        \caption{The spectra of GRB\,180128A over four BB time intervals. The intervals show the onset of the burst (1), peak 1 (2), peak 2 (4), and the extended emission after the peaks (6). Intervals (3) and (5) are omitted for clarity yet still consistent with the trend displayed. The shaded area indicates the $1\sigma$ confidence regions}
        \label{fig:GRB180128A_nuFnu}
    \end{minipage}
\end{figure}

\begin{deluxetable*}{cccccc}
\tabletypesize{\footnotesize}
\tablecaption{\textbf{Time-resolved spectral analysis using Bayesian Blocks}}
\tablewidth{0pt}
\tablehead{
\colhead{Time} & \colhead{$E_{\mathrm{p}}$} & \colhead{$\alpha$} &  \colhead{Energy Flux ($\cal{F}$)} & \colhead{$L_{\mathrm{iso}}$} &  \colhead{$E_{\mathrm{iso}}$}\\
\colhead{(ms)} & \colhead{(keV)} & \colhead{} &  \colhead{(x10$^{-6}$)~\fluxcgs}  & \colhead{x10$^{47}$~erg$\cdot$~s$^{-1}$} &  \colhead{x10$^{45}$~erg}} 
\startdata
\multicolumn{6}{c}{\textbf{GRB\,180128A}}\\
-12:-10 & 250 $\pm$ 50 & 6.0 $\pm$ 4.9 & 17.1 $\pm$ 4.6 & 0.28 $\pm$ 0.08 & 0.06 $\pm$ 0.02\\
-10:-7 \textbf{(Peak 1)} & 560 $\pm$ 140 & 0.7 $\pm$ 0.6 & 64.5 $\pm$ 9.5 & 1.1 $\pm$ 0.2  & 0.32 $\pm$ 0.05\\
-7:-3 & 400 $\pm$ 120 & 8.2 $\pm$ 13.2 & 10.3 $\pm$ 4.1 & 0.17 $\pm$ 0.07 & 0.07 $\pm$ 0.03\\
-3:-1 \textbf{(Peak 2)} & 490 $\pm$ 150 & 0.7 $\pm$ 0.8 & 50 $\pm$ 10 &  0.75 $\pm$ 0.17 & 0.15 $\pm$ 0.03\\
-1:18 & 190 $\pm$ 20 & 4.9 $\pm$ 2.3 & 3.7 $\pm$ 0.7 & 0.060 $\pm$ 0.010  & 0.12 $\pm$ 0.02\\
18:143 & 120 $\pm$ 30 & 1.6 $\pm$ 1.7 &  0.40 $\pm$ 0.11 &  0.0070 $\pm$ 0.0002  & 0.081 $\pm$ 0.002\\
\hline
\textbf{T$_{BB}$ duration (155):} & 290 $\pm$ 50 & 0.6 $\pm$ 0.5 & 2.4 $\pm$ 0.4 & 0.039 $\pm$ 0.007 & 0.60 $\pm$ 0.10\\
\hline\hline
\multicolumn{6}{c}{\textbf{GRB\,200415A}}\\
-4.3:-3.9 \textbf{(Peak 1)} & 320 $\pm$ 50 & 0.5 $\pm$ 0.4 & 170 $\pm$ 20 & 2.7 $\pm$ 0.4 & 0.11 $\pm$ 0.02\\
-3.9:-3.4 & 1100 $\pm$ 700 & -0.8 $\pm$ 0.3 & 120 $\pm$ 60 & 2.0 $\pm$ 0.9 & 0.10 $\pm$ 0.05\\
-3.4:-2.9  \textbf{(Peak 2)}& 800 $\pm$ 130 & -0.1 $\pm$ 0.2 & 490 $\pm$ 80 & 8.0 $\pm$ 1.3 & 0.40 $\pm$ 0.07\\
-2.9:-2.5  \textbf{(Peak 3)}& 1100 $\pm$ 200 & -0.50 $\pm$ 0.14 & 850 $\pm$ 140 & 14 $\pm$ 2 & 0.56 $\pm$ 0.10\\
-2.5:-0.5$^{\dagger}$ & 1210 $\pm$ 120 & -0.1 $\pm$ 0.1 & 630 $\pm$ 60 & 10.5 $\pm$ 1.0 & 2.1 $\pm$ 0.2\\
-0.5:3.0 & 2040 $\pm$ 180 & -0.16 $\pm$ 0.08 & 600 $\pm$ 50 & 10.0 $\pm$ 1.0 & 3.4 $\pm$ 0.3\\
3.0:5.0$^{*}$ & 900 $\pm$ 200 & 0.1 $\pm$ 0.3 & 110 $\pm$ 30 & 1.8 $\pm$ 0.4 & 0.36 $\pm$ 0.07\\
5.0:6.5$^{*}$ & \multicolumn{5}{c}{Completely in the data gap} \\
6.5:22.5$^{*}$ & 1040 $\pm$ 80 & 0.8 $\pm$ 0.2 & 124 $\pm$ 9 & 2.0 $\pm$ 0.2 & 3.3 $\pm$ 0.2\\
22.5:65.8 & 830 $\pm$ 50 & 0.46 $\pm$ 0.13 & 51 $\pm$ 3 & 0.84 $\pm$ 0.05 & 3.7 $\pm$ 0.2\\
65.8:93.4 & 590 $\pm$ 70 & 0.3 $\pm$ 0.2 & 19 $\pm$ 2 & 0.31 $\pm$ 0.03 & 0.85 $\pm$ 0.09\\
93.4:121.2 & 430 $\pm$ 50 & 0.8 $\pm$ 0.4 & 9.7 $\pm$ 1.1 & 0.16 $\pm$ 0.02 & 0.44 $\pm$ 0.05 \\
121.2:150.3 & 250 $\pm$ 40 & 0.9 $\pm$ 0.6 & 3.1 $\pm$ 0.5 & 0.051 $\pm$ 0.007 & 0.15 $\pm$ 0.02\\
\hline
\textbf{T$_{BB}$ duration (155)$^{\dagger\dagger}$:} & 998 $\pm$ 40 & 0.04 $\pm$ 0.05 & 56 $\pm$ 2 & 0.91 $\pm$ 0.04 & 14.2 $\pm$ 0.5\\
\hline\hline 
\multicolumn{6}{c}{\textbf{GRB\,150101B}}\\
-16:-8 \textbf{(Peak 1)} & 1100 $\pm$ 600 & -0.4 $\pm$ 0.4 & 16 $\pm$ 7 & 8000 $\pm$ 4000 & 6000 $\pm$ 3000\\
-8:-2 \textbf{(Peak 1)} & 220 $\pm$ 70 & -0.8 $\pm$ 0.3 & 6.3 $\pm$ 1.3 & 3200 $\pm$ 700 & 1900 $\pm$ 400\\
-2:4 & 61 $\pm$ 12 & 0.3 $\pm$ 1.2 & 1.4 $\pm$ 0.3 & 730 $\pm$ 150 & 440 $\pm$ 90\\
4:74 & 25 $\pm$ 6 & -0.8 $\pm$ 1.0 & 0.35$\pm$ 0.05 & 180 $\pm$ 30 & 1300 $\pm$ 200\\
\hline
\textbf{T$_{BB}$ duration (90):} & 150 $\pm$ 80 & -1.5 $\pm$ 0.2 & 1.1$\pm$ 0.2 & 550 $\pm$ 120 & 4100 $\pm$ 900\\
\enddata
\tablecomments{The fluence is from fitting the spectrum with a Comptonized function over a combined (NaI and BGO detectors) spectral range of 8 keV to 40 MeV.\\
$^{\dagger}$ Includes the saturated portion of the spectrum from about T$_{0}$-2.4 to -0.8~ms. \\
$^{*}$ Includes the data gap, from about T$_{0}+4.6$ to 6.6~ms.\\
$^{\dagger\dagger}$ This does not include the correction for the brightest part of the event that was saturated in \fermi\,as in \cite{2021Natur.589..207R}, which accounts for the lower values in this study.}
\end{deluxetable*}\label{tab:GBM_spec}

\subsection{\textit{Fermi}~GBM Analysis of GRB~180128A}
 \label{sec:gbmtempanal}

GRB\,180128A triggered GBM on January 28$^{th}$, 2018 at 05:09:56.60 UT. We generate responses for detectors viewing that position within 60$^{\circ}$ of their boresight.
The initial $T_{50}$ and $T_{90}$ durations of GRB\,180128A were found to be 48 $\pm$ 51~ms and 208 $\pm$ 400~ms, respectively~\citep{von2020fourth}. These error bars are typical for short-duration bursts where the total fluence is almost comparable to background fluctuations in GBM. Reanalyzing this event using BB analytical techniques~\citep{scargle2013bayesian}, we find a duration ($T_{BB}$) of 155~ms. 

We initially fit the differential energy spectrum of GRB\,180128A with models typically employed for GRBs~\citep{von2020fourth}, and also MGFs~\citep{2021Natur.589..207R,svinkin2021bright,kaspi2017}. These fits included a simple power-law (PL) model and a Band function~\citep{1993ApJ...413..281B}. In addition, our spectral fitting models focused on a Comptonized function, COMPT \citep{gruber2014fermi}, a form common to magnetar burst studies \citep[e.g.,][]{Lin-2011-ApJ}. The function is a power-law of index $\alpha$ modulated by an exponential cutoff at a characteristic energy $E_{\rm p}$ that is close to the spectral peak in a $\nu F_{\nu}$ representation. The values of $\alpha$ and $E_{\rm p}$ for GRB\,180128A, and the two comparison bursts are listed in Table~\ref{tab:GBM_spec} for an array of BB time bins, defining the spectral evolution of these transients. To better illustrate the soft-to-hard-to-soft evolution that is the envelope behavior of this burst, we plot four of the BB intervals of GRB\,180128A. The results in Figure~\ref{fig:GRB180128A_nuFnu} exhibit the spectral evolution from the onset of the burst to the two peaks to the extended emission after the peaks. For clarity, we omit the third and fifth BB intervals from Figure\,\ref{fig:GRB180128A_nuFnu}. They are, however, consistent with the trend seen in the displayed intervals. Notably, this spectral evolution is reminiscent of that clearly evident in the MGF GRB\,200415A.

We perform a time-integrated fit using a COMPT model to the \fermi\,data for GRB\,180128A over energies 8\,keV--40\,MeV, and the beginning and end of the significant emission as defined by the BB analysis, the results of which are in Table\,\ref{tab:GBM_spec}. We note that a Band model fits the spectrum equally well but with more poorly constrained values. Using the distance to NGC~253 of 3.7~Mpc \citep[][at this distance, we neglect cosmological redshift]{leroy2019z}, we find an $E_{\mathrm{iso}}$ = (5.9 $\pm$ 1.0)$\times$10$^{44}$\,erg from the fluence values of the COMPT fit to the spectrum over the BB duration of 155\,s. We also find an $L_{\mathrm{iso}}$ = (3.9 $\pm$ 0.7)$\times$10$^{45}$\,erg s$^{-1}$ from the fluence and flux values for the same model fit and duration.

We first compared the spectroscopy of our burst with that for the MGF GRB\,200415A \citep{svinkin2021bright,2021Natur.589..207R}, the only other MGF clearly identified by \fermi\, up to now, fitting the time bins shown in the lightcurve in Figure\,\ref{fig:GBM_LC}. The $L_{\mathrm{iso}}$ and $E_{\mathrm{iso}}$ determinations for GRB\,180128A and GRB\,200415A were consistent with those expected from a MGF and are given in Table~\ref{tab:GBM_spec} for each BB time interval, along with those for GRB\,150101B. For GRB\,200415A and GRB\,150101B, the values from our time-integrated COMPT model analysis (also listed in Table~\ref{tab:GBM_spec}) were consistent with those obtained in~\cite{2021Natur.589..207R} and \citet{2018ApJ...863L..34B}, respectively. It is worth noting that the $E_{\rm{iso}}$, $E_{\rm{p}}$, and $\alpha$ values list for GRB\,180128A and GRB\,200415A fall within the known values of all other known MGFs \citep{burns2021identification}: with $\alpha$ ranging from around 0.0 to 1.0 and $E_{\rm{p}}$ starting around 300\,keV and extending as high as several MeV.

The detection of GRB\,150101B by \textit{Swift} BAT localized its position well \citep{2015GCN.17267....1C}, so that follow-up observations determined a host galaxy redshift $z = 0.134$~\citep{Fong2016}, corresponding to a source distance of 0.65 Gpc from Earth using the most recent Planck Collaboration $\Lambda$CDM cosmology parameters~\citep{Planck-2020-AandA}. This distance is over two orders of magnitude greater than that of NGC\,253, implying that this event is several orders of magnitude more luminous than the known Galactic MGFs. We chose GRB\,150101B for comparison because it is one of the seven candidates that met all initial temporal selection criteria, but the known distance excludes a MGF origin. This transient is likely the result of a NS merger, analogous to GRB\,170817A~\citep{2017ApJ...848L..14G,2018NatCo...9.4089T,2018ApJ...863L..34B}.

The lightcurve of GRB\,150101B consists of a 16~ms spike followed by a longer, softer decaying tail that lasts around 64~ms, for a total event duration of around 80~ms~\citep{2018ApJ...863L..34B}, and is generated using the same criterion for determining detectors with good angles relative to the source (i.e., angle $<60^{\circ}$ to boresight). Given the cosmological distance, we can not neglect cosmological redshift. Therefore, to calculate the  $L_{\mathrm{iso}}$ and $E_{\mathrm{iso}}$ of the time-resolved and time-integrated spectra, we must k-correct the energy measurements \citep{bloom2001prompt}.

\subsection{{Fermi}~LAT Analysis of GRB~180128A}
\label{sec:LATana}

The \textit{Fermi}\,LAT \citep{theLAT} detected high-energy emission from the NGC~253 MGF in April 2020 \citep[][GRB\,200415A]{LATMGF}. The detection consisted of three photons between 0.48 and 1.7 GeV. The arrival time of the first photon was delayed with respect to the GBM trigger by $\sim$19\,s, with the last photon detected $\sim$284\,s later. With the tight sky localization of these photons compatible with the location of NGC~253 and the approximate temporal coincidence with the MGF, this data represented the first ever \textit{Fermi}\,LAT detection in the \gev energy band of emission from a magnetar.

For GRB\,180128A, NGC~253 was well within the LAT field of view at the moment of the trigger and remained visible until $\sim$3000 s after. We find no significant detection at the source location and place an upper limit on the source flux. We performed a likelihood analysis with {\tt gtburst}\footnote{\url{https://fermi.gsfc.nasa.gov/ssc/data/analysis/scitools/gtburst.html}}, using P8R3 data with TRANSIENT10e source class, and select events within a 5-degree radius from the target, with energy between 100 MeV and 10 GeV in a time window between 0 and 3000 s from the trigger time. Assuming a power law spectrum with an index $\Gamma=-2.1$, we find an energy flux upper limit of $2.3\times10^{-10}$ erg cm$^{-2}$ s$^{-1}$. Considering an integration time of 3000\,sec and the distance to NGC~253 of 3.7 Mpc, the limit on the intrinsic total energy above the LAT threshold is $E_{{\rm iso}} < 1.8\times 10^{44}$\,erg at a 95\% confidence level. The MGF detected by the LAT in April 2020 had an estimated intrinsic energy of $(3.6\pm2.1)\times10^{45}$\,erg in the 0.1--10 GeV band, almost one order of magnitude higher than the upper limit we find in this work. The non-detection of GRB\,180128A clearly rules out any notion that all MGFs have a GeV counterpart of similar strength. 

Yet, we cannot {\it a priori} exclude that there is a quasi-linear scaling of the GeV emission observed by the LAT with the soft gamma-ray luminosity detected by the GBM. 

To assess this possibility and consistently compare GRB\,180128A with GRB\,200415A, we repeat the analysis routine with the same setup as in \cite{LATMGF}, namely with {\tt P8\_TRANSIENT020E} events in a time window 10$-$500 s from the GBM trigger. This time, we select a region of 12 degrees radius to match the analysis setup of GRB\,200415A chosen to accumulate enough background statistics to allow a better fit of the Galactic diffuse and isotropic emissions. According to this setup, the upper limit at 95\% C.L. on the intrinsic luminosity in the 0.1-10 GeV energy range is $L_{{\rm iso}} < 2.3\times 10^{42}$ erg/s ($E_{\rm iso} < 1.1\times 10^{45}\,\rm{erg}$), more than a factor of 11 above the predicted value given by the simple scaling of the intrinsic energetic luminosity found for GRB\,200415A. 
The upper limits we obtain using this setup are significantly larger, by an order of magnitude, compared to the limits for GRB\,180128A obtained with the previous setup and listed in Table\,\ref{tab:LAT}. These new limits are much closer to the values listed in Table\,\ref{tab:LAT} for GRB\,200415A.

\begin{deluxetable*}{lcccc}[ht]
\tabletypesize{\footnotesize}
\tablecaption{\textbf{\textit{Fermi}-LAT results summary}}
\tablewidth{0pt}
\tablehead{
\colhead{Name} & \colhead{Flux} & \colhead{Energy Flux} &  \colhead{$L_{\mathrm{iso}}$} & \colhead{$E_{\mathrm{iso}}$} \\
\colhead{} & \colhead{(cm$^{-1}$s$^{-1}$)} & \colhead{(erg cm$^{-1}$s$^{-1}$)} &  \colhead{(erg s$^{-1}$)} & \colhead{(erg)}}
\startdata
GRB\,180128A & $< 1.9\times10^{-6}$ &  $< 2.3\times10^{-10}$ & $< 3.7\times10^{41}$ & $< 1.8\times10^{44}$ \\
GRB\,200415A &  $(4.1\pm2.2)\times10^{-6}$ & $(4.8\pm2.7)\times10^{-9}$ & $(7.4\pm4.2)\times10^{42}$ & $(3.6\pm2.1)\times10^{45}$ \\
GRB\,150101B &  $< 4.9\times10^{-6}$ & $< 1.9\times10^{-9}$ & $< 9.4\times10^{46}$ & $< 2.8\times10^{50}$ \\
\enddata
\tablecomments{Comparison between the LAT detection of GRB\,200415A and the upper limit on GRB\,180128A and GRB\,150101B. For the latter, we assume z=0.134 for the associated host galaxy ($D=0.65$ Gpc) \cite{Fong2016}, and we use the same analysis setup as for GRB\,180128A (see text). All fluxes and energy fluxes measured from 0.1 to 10 GeV.}
\end{deluxetable*}\label{tab:LAT}

\subsection{Optical}
\label{sec:Optical}

To rule out the possibility that GRB\,180128A was due to other transient candidates visible in the optical band, we conduct searches of the data from the Zwicky Transient Facility (ZTF) \citep{2019PASP..131g8001G, 2019PASP..131a8002B} and MASTER \citep{lipunov2010master, 2022Univ....8..271L}. ZTF data show no obvious candidates, ruling out a SN and the typical afterglow at the distance of NGC~253. However, there are no observations between December $12^{th}$, 2017 and July $22^{nd}$, 2018, the range which overlaps the trigger time of our event. The MASTER observations covered $\sim 3/4$ of the IPN localization up to February 4th ($\sim 8$ days after the burst) with no optical transient detections down to $\sim 17$--19 mag, which suggests that the burst was likely not associated with a SN.

\subsection{Gravitational waves}
\label{sec:GW} 

The Laser Interferometer Gravitational-Wave Observatory (LIGO)~\citep{LIGOScientific:2014pky} and Advanced Virgo~\citep{AdvancedVirgo} were offline at the time of the event, having ended the second observing run on August 25$^{th}$, 2017\footnote{\url{https://www.ligo.caltech.edu/page/timeline}}. Hence, there are no gravitational-wave (GW) data to determine whether this event might be due to a NS merger.

LIGO and Virgo also have searches targeting GWs produced from magnetars~\citep{LIGOScientific:2010jrd,KAGRA:2021bhs}, including those associated with MGFs~\citep{LIGOScientific:2019ccu,Macquet:2021eyn}. Even in the absence of direct detection, it is possible to set upper limits on the GW energy emission. In fact, during the Galactic MGF of December $27^{th}$ 2004 \citep{palmer2005giant}, LIGO reported an upper limit on the energy emission \citep{abbott2007search}. However, since this initial reporting, the sensitivity of LIGO detectors has increased $\sim$100x. Given this increase in sensitivity, the LIGO-Virgo-KAGRA observations from future observing runs~\citep{abbott:hal-02973168} will provide important information for describing the energetics of identified MGFs.

\subsection{Radio}
\label{sec:radio}

Associations between fast radio bursts (FRBs) and coincident X-ray bursts from magnetars \citep{bochenek2020fast} have led to the development of models for these phenomena consistent with MGFs. Despite NGC\,253 being ``radio-loud'' due to a prominent synchrotron radio halo \citep{1992ApJ...399L..59C}, to determine if there were any FRBs detected coincident with GRB\,180128A, we consult the FRB\footnote{\url{https://www.frbcat.org/}} \citep{2016PASA...33...45P} and CHIME\footnote{\url{https://www.chime-frb.ca/catalog}} \citep{2021ApJS..257...59C} catalogs. The FRB catalog lists two non-verified events on the same day as GRB\,180128A. However, both detection localizations are far outside the IPN localization for our event. No relevant events listed in the CHIME catalog coincide with the time or location of GRB\,180128A. A search of the Very Large Array archive \citep{perley2011expanded} also found no observations around the time of our event.

\section{Discussion}
\label{sec:GBMDisInc}

Many of the characteristics seen in GRB\,180128A can be explained by current MGF theory. However, it displays several properties that make it a unique event. The first being the appearance of two distinct peaks in the lightcurve. The other interesting point is that this burst was localized to the same galaxy as another known MGF, making it the first time multiple MGFs are associated with the same galaxy outside the local neighborhood. Also of interest is how different binning techniques affect how the relativistic wind structures are resolved. Here, we discuss the implications of these properties.

\subsection{Physical Mechanisms and Spectral-Flux Correlations in \textit{Fermi}~GBM Data for the Giant Flare Candidates}
\label{sec:gbm-Liso-Epeak}

The energetics analysis in Section\,\ref{sec:GBManalaysis} can be interpreted as a result of a large-scale crustal disruption event \citep{Norris1991ApJ...366..240N,1995MNRAS.275..255T}. Due to a build-up of magnetic stresses in the crust, the stellar crust reaches a breaking point and can shear and crack, rapidly heating the local plasma. This powerful event launches a hot, relativistic pair/photon fireball with little baryonic contamination into the magnetosphere (\citep{1995MNRAS.275..255T}, accompanied by magnetic reconnection as the plasmoid is ejected. The fast initial pulse of GRB180128A is consistent with this very large release of magnetic energy \citep{1995MNRAS.275..255T, 1996ApJ...473..322T}.

\begin{figure}[ht]
    \centering
    \begin{minipage}[t]{0.9\textwidth}
        \centering
        \includegraphics[width=\linewidth]{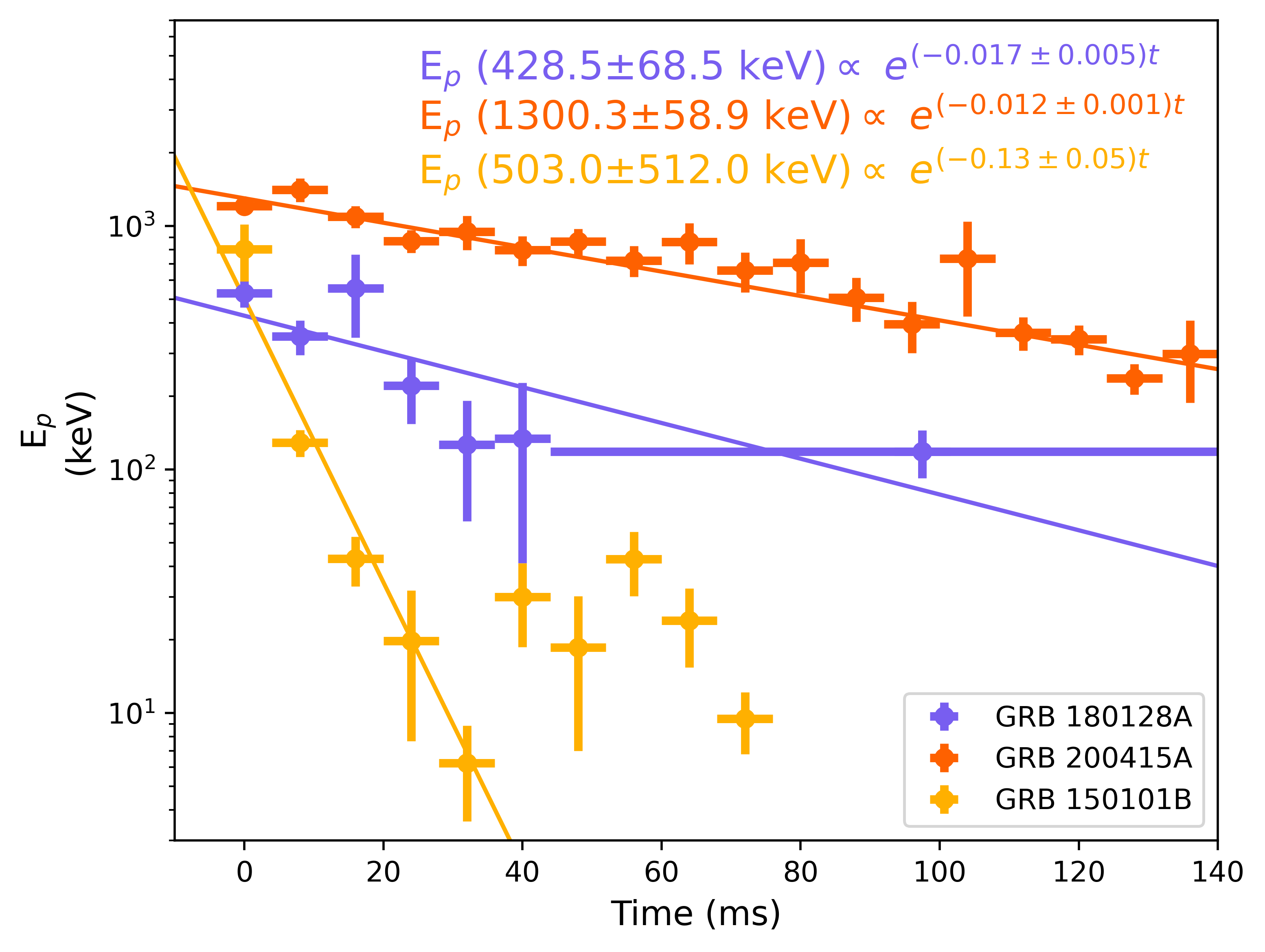}
        \caption{The Comptonized spectrum peak energy ($E_{\mathrm{p}}$) as a function of time using a temporal binning of 8\,ms. All fit errors and error bars are at the 1-$\sigma$ confidence level. The zero-time reflects the GBM event start time of each GRB.}
        \label{fig:GBM_Ep_time}
    \end{minipage}
\end{figure}

\begin{figure}[hbt!]
    \begin{minipage}{.5\textwidth}
        \begin{center}
        \includegraphics[width=\linewidth]{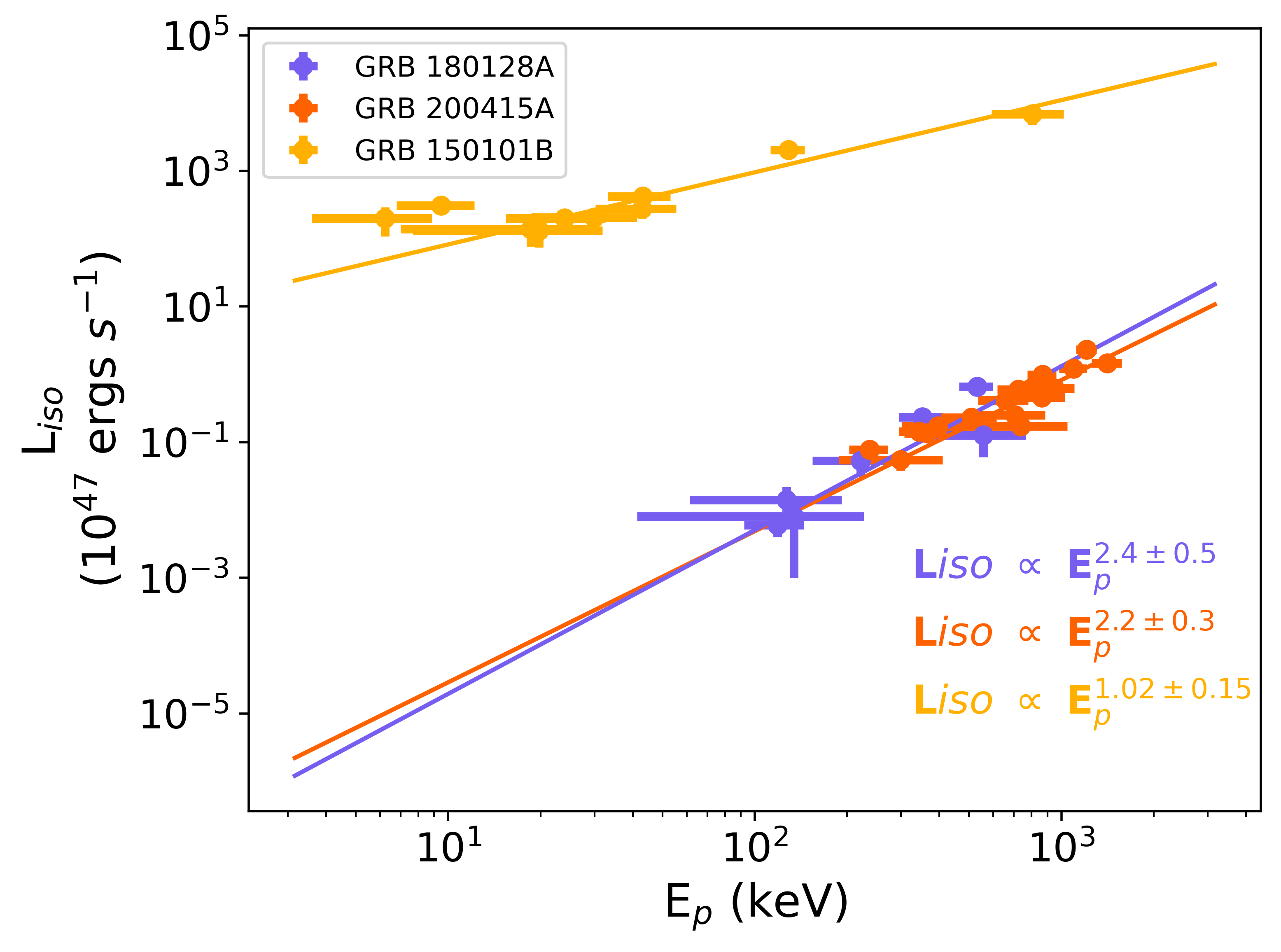}\\
        \textbf{(a)}
        \end{center}
    \end{minipage}
    \begin{minipage}{.5\textwidth}
        \begin{center}
        \includegraphics[width=\linewidth]{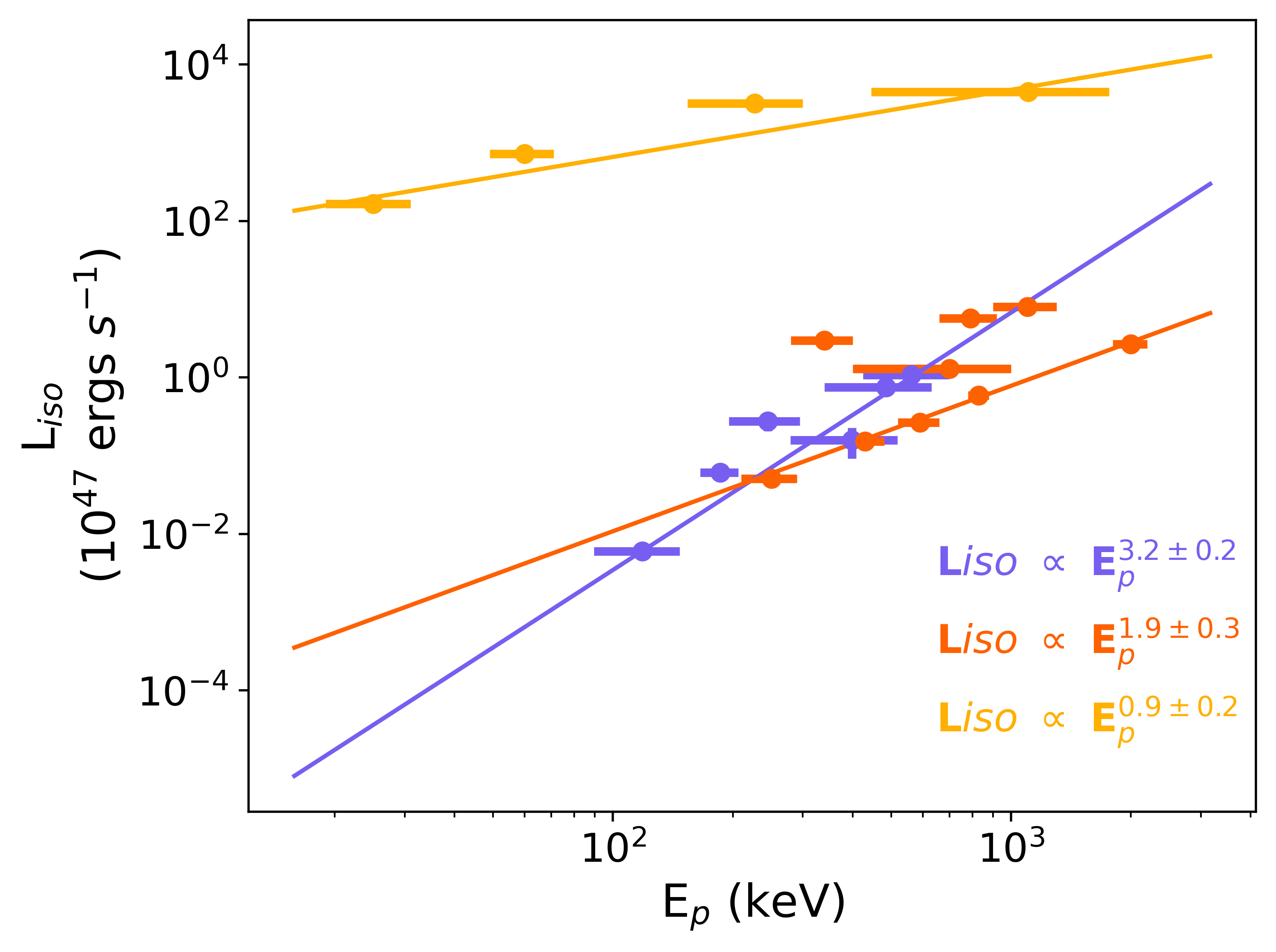}\\
        \textbf{(b)}
        \end{center}
    \end{minipage}\hfill 
    \caption{Flux and spectral evolution of GRB\,180128A (lavender), GRB\,200415A (orange), and GRB\,150101B (amber), color-coded as in the inset in panel (a). \textbf{Left (a):} The correlation between $L_{\mathrm{iso}}$ and $E_{\mathrm{p}}$ for all three transients, revealing an approximate $L_{\mathrm{iso}}$ $\propto$ $E_{\mathrm{p}}^{2}$ relationship that is a strong signature of relativistic winds. The temporal binning for panels a is uniformly 8~ms. \textbf{Right (b):} The $L_{\mathrm{iso}}$ and $E_{\mathrm{p}}$ for all three transients over the BB intervals in Table\,\ref{tab:GBM_spec}, omitting the data gaps and saturated intervals of GRB\,200415A. All fit errors and error bars are at the 1-$\sigma$ confidence level.}
    \label{fig:GBM_Ep_evol_Liso}
\end{figure}

The classification of GRB\,180128A as a MGF came from the initial selection criteria and localization. The analysis shows it is consistent with a MGF origin. Further, insights into the MGF physical emission region and mechanism originate from considerations of the spectral evolution and the coupling between spectral and flux variations. We focus on the $\nu F_{\nu}$ peak energy $E_{\rm p}$. For the bright GRB\,200415A, the values for the power law index, $\alpha$, in the COMPT spectral fits are generally in the range $0-1$ (see Table~\,\ref{tab:GBM_spec}). This is commensurate with expectations from polar winds in MGFs due to the high opacity of Compton scattering by electrons in the strong magnetic fields \citep{2021Natur.589..207R}. For our spectral-flux correlation analysis of GRB\,200415A, we employed the $\alpha$ values determined from the COMPT fit, which match those of \citet{2021Natur.589..207R} for Figure~\ref{fig:GBM_Ep_time} and Figure~\ref{fig:GBM_Ep_evol_Liso}a. In contrast, because the count statistics for GRB\,180128A and GRB\,150101B was poorer, we fixed the power-law index to $\alpha=1.0$ and $\alpha=0.0$, respectively, for Figure~\ref{fig:GBM_Ep_time} and Figure~\ref{fig:GBM_Ep_evol_Liso}a. These values assume both bursts have a MGF origin: GRB\,180128A being lower energy, we expect an $\alpha$ closer to 1.0, while GRB\,150101B has a much higher energy, which would lead to an expected $\alpha$ value closer to 0.0. Any deviation from behaviors expected from a MGF using these values would strongly disfavor a likely MGF origin. For all three bursts in Figure~\ref{fig:GBM_Ep_evol_Liso}b, the adopted $\alpha$ values are those listed for the BB interval choice in Table~\ref{tab:GBM_spec}.

Figure~\ref{fig:GBM_Ep_time} plots the evolution of $E_{\rm p}$ values for our two selected MGF candidates plus GRB\,150101B, with data for each burst acquired using 8\,ms temporal binning. It displays a quasi-exponential decay for GRB\,200415A on a relatively long timescale \citep[see also][]{2021Natur.589..207R}. GRB\,180128A has a more rapid decay that is quasi-exponential at its outset. In contrast, GRB\,150101B exhibits $E_{\rm p}$ values with a faster decay that quickly morphs into an $E_{\rm p}$ fluctuation.

Figure~\ref{fig:GBM_Ep_evol_Liso}a presents the relationship between $L_{\mathrm{iso}}$ and $E_{\mathrm{p}}$, also temporally binned to 8\,ms intervals. For GRBs\,1801128A and 200415A, we see a well-defined $L_{\mathrm{iso}}$ $\propto$ $E_{\mathrm{p}}^{2}$ relationship which is a strong indicator of relativistic Doppler boosting \citep[][see also just below]{2021Natur.589..207R}. The $L_{\mathrm{iso}}\propto\,E_{\mathrm{p}}^{2}$ relation is not as readily recovered for GRB\,150101B, which has fewer data points. Interestingly, when plotting the $L_{\rm iso}$-$E_{\rm p}$ correlation using the BB intervals (omitting any data gaps or saturated data points) as in Figure\,\ref{fig:GBM_Ep_evol_Liso}b, the fit index is no longer 2 (i.e., $L_{\mathrm{iso}}$ is not $\propto\,E_{\mathrm{p}}^{2}$), and differs between the MGF candidates. The fit index for GRB\,180128A is greater than 3, yet there is significant dispersion in the data points that closely matches the scatter in the time-dependent analysis of \cite{chand2021magnetar}. Moreover, we observe that the $L_{\mathrm{iso}}\propto\,E_{\mathrm{p}}^{2}$ relation can be recovered when fitting the lower $L_{\rm iso}$ intervals that correspond to the dips in the lightcurve of GRB\,200415A. The four points above the fit line for GRB\,200415A displayed in Figure\,\ref{fig:GBM_Ep_evol_Liso}b correspond to the first four BB intervals (and thus the peaks in the lightcurve) in Table~\ref{tab:GBM_spec}. This selective sampling suggests that the BB binning choice resolves the finer structures within the burst.

The dependence of the $L_{\mathrm{iso}}$-$E_{\mathrm{p}}$ correlation on the temporal binning protocol is not unexpected, and it provides interesting insight into the observational sampling of a MGF wind. \cite{2021Natur.589..207R} interpreted the $L_{\mathrm{iso}} \propto\, E_{\mathrm{p}}^{2}$ coupling for GRB\,200415A as being the signature of a relativistic wind emanating from the magnetar pole and collimated by the field lines (essentially a flared jet) that sweeps across an observer's line of sight as the star rotates. During this sweep, the effective Doppler factor $\delta_{\rm w}$ sampled by {\it Fermi} GBM rises and falls, generating a spread in $E_{\mathrm{p}}$, with $E_{\mathrm{p}} \propto \delta_{\rm w}$ for the photon energy Doppler blueshift, and a range in $L_{\mathrm{iso}}$. For the temporally agnostic choice of fixed, 8\,ms time bins, the cumulative flux in a bin tends to sample a broader range of wind axis orientations during the stellar rotation. From the perspective of the jet axis, this essentially integrates over a large solid angle $\Delta \Omega$ of the Doppler beam. Accordingly, the accumulated signal is a flux rather than an intensity and depends quadratically on the Doppler factor, i.e., $L_{\mathrm{iso}} \propto \delta_{\rm w}^2$ due to the combination of photon Doppler blueshift and time dilation in the Lorentz boost from the wind frame to that of the observer. Thus the $L_{\mathrm{iso}}$ $\propto$ $E_{\mathrm{p}}^{2}$ relationship naturally emerges, as highlighted in the analysis of \cite{2021Natur.589..207R} and Figure\,\ref{fig:GBM_Ep_evol_Liso}a for GRB\,200415A, and also for GRB\,1801128A. 

In contrast, the BB approach inherently bins on shorter timescales for higher fluxes (see Figure\,\ref{fig:GBM_LC}), corresponding to a narrower sampling of solid angles relative to the jet axis as the magnetar rotates. Then the detected signal scales more like intensity, i.e., flux per unit solid angle, and this has the well-known Doppler boosting dependence $L_{\mathrm{iso}} \propto \delta_{\rm w}^4$ since $\Delta \Omega\propto \delta_{\rm w}^{-2}$ defines the Doppler cone. Accordingly, one would then anticipate a $L_{\mathrm{iso}}$ $\propto$ $E_{\mathrm{p}}^{4}$ in this particular extreme. It is then no surprise that the BB display in Figure\,\ref{fig:GBM_Ep_evol_Liso}b for GRB\,200415A and GRB\,180128A generates a stronger $L_{\mathrm{iso}}$-$E_{\mathrm{p}}$ correlation than that observed for fixed time bins in Figure\,\ref{fig:GBM_Ep_evol_Liso}a. The fit index being below four suggests that there is a partial sampling of the wings of the Doppler beam in the shortest BB time bins, which, in principle, possibly allows for an estimate of the ratio of the unknown rotation period to the bulk Lorentz factor $\Gamma  \sim \delta_{\rm w}$ of the MGF wind. Observe that the flat $L_{\mathrm{iso}}$-$E_{\mathrm{p}}$ correlation for GRB\,150101B cannot be explained by this picture. Its poorer count statistics and anomalous light curve suggest that other conditions prevail: its wind may not be ultra-relativistic, thereby limiting its Doppler beaming, with both being consistent with GRB\,150101B resulting from a NS merger as opposed to being a MGF.

\begin{figure}[ht]
    \centering
    \begin{minipage}[t]{0.75\textwidth}
        \centering
        \includegraphics[width=\linewidth]{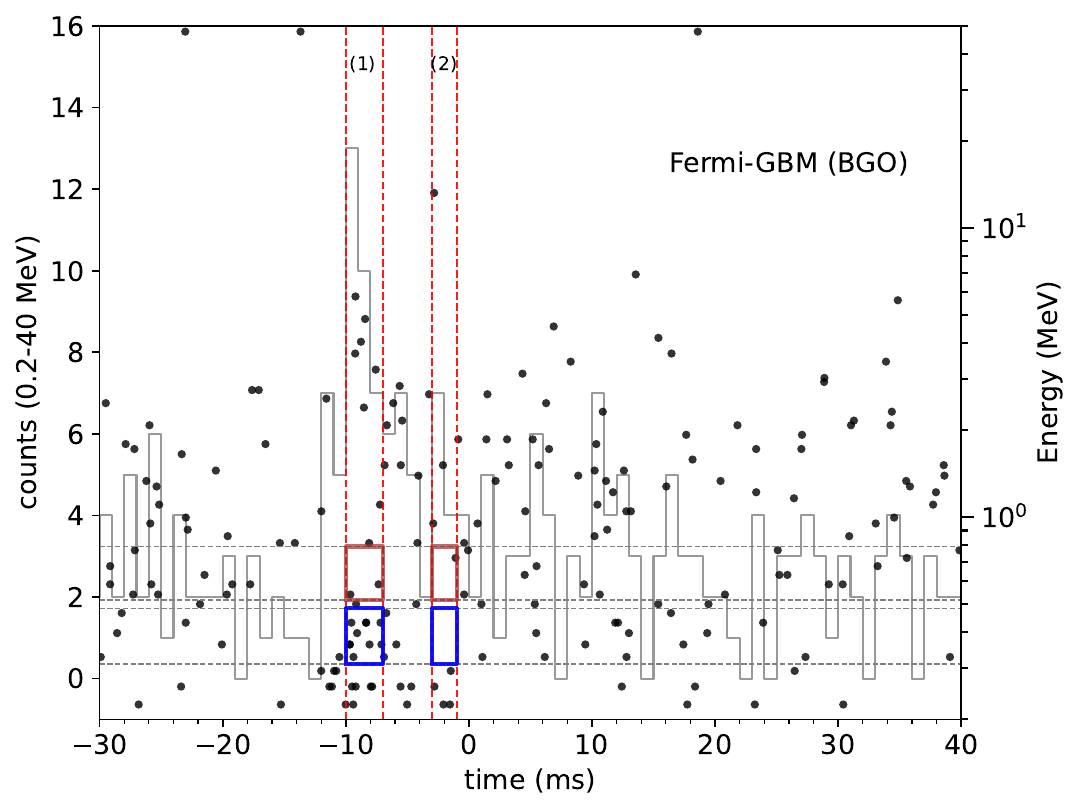}
        \caption{Individual TTE data of GBM BGO detector 1 (black dots) for GRB\,180128A. The blue rectangles indicate energies from 312 to 484 keV in intervals (1) and (2), corresponding to peaks 1 and 2. The red rectangles represent energies from 518 to 792 keV for the same time intervals. We conclude that the highest photon energy associated with GRB\,180128A is $\sim 500 \rm{keV}$.}
        \label{fig:TTE_counts}
    \end{minipage}
\end{figure}

To address the element of how fast the MGF winds are, we follow the time-honored GRB tradition of using the argument of transparency of the highest energy photon (of energy $E_{\rm max}$) to QED pair creation $\gamma\gamma \to e^+e^-$ to bound the bulk Lorentz factor $\Gamma$ of the wind. All detected photons must escape from the emission region, and relativistic Doppler beaming makes this easier as collimation of the photons increases the effective pair threshold energy \citep{Krolik-1991-ApJ,Baring-1993-ApJ}. The most conservative bound of $\Gamma > E_{\rm max}/511\,$keV is obtained using Doppler boosting of the 511 keV pair threshold from the wind rest frame. For the GRB\,200415A MGF, \cite{2021Natur.589..207R} reported $E_{\rm max}\sim 3\,$MeV in the GBM data (about a factor of 3 higher than the typical $E_{\rm p}$ listed in Table~\ref{tab:GBM_spec}) and deduced $\Gamma \gtrsim 6$, though this is well below that inferred from the \textit{Fermi}-LAT observations of delayed GeV photons \citep[][see just below]{LATMGF}. Figure\,\ref{fig:TTE_counts} shows the individual counts from the TTE data of GRB\,180128A for the \fermi\,BGO 1 detector, with regions (1) and (2) corresponding to the two peak intervals. Using the Bayesian method described in \cite{2021Natur.589..207R} for the on-source/off-source signal detection, a calculated probability of 0.9999896 that an excess signal above background in the 312--484\,keV region of peak 1 (blue box in interval 1) can be attributed to GRB\,180128A. We find no statistically significant excess source signals for the remaining energy regions. The highest energy of photons from GRB\,1801128A identified as being statistically significant above background is $\sim 500\,$keV, implying a bound $\Gamma \gtrsim 1$ that is clearly much less relativistic than that for GRB\,200415A. Thus, pair transparency bounds on $\Gamma$ are poorly defined for GRB\,180128A.

\subsection{Multi-Pulse Variability}
\label{pulse_variability}

The lightcurve of GRB\,180128A displays two distinct pulses at sub-ms resolution (Figure\,\ref{fig:GBM_LC}). This MGF candidate is the fourth event to display such behavior, the other three being GRB\,070201, \citep{mazets2008giant, 2008ApJ...681.1464O}, GRB\,070222 \citep{burns2021identification}, and GRB\,200415A \citep{2021Natur.589..207R}. In the case of GRB\,070201, the temporal variability was used to argue against it having a MGF origin and raised the possibility that it was a background short GRB \citep{2008ApJ...681.1464O}, as such variability is consistent with short GRBs \citep{nakar2002temporal}. These four account for roughly half of all likely events, the majority of which do not have sub-ms data, which may point to this being a common characteristic of MGFs.

The multi-pulse variability may imply repeated injections or varying observer geometry with the outflow. From Figure~\ref{fig:GBM_Ep_evol_Liso}a, it is apparent that GRB\,180128A’s interpulse traces $d\log{L_{\mathrm{iso}}}/(d\log E_{\mathrm{p}}) \sim 2$ as for other MGF candidates (and is distinct from GRB\,150101B). This relation is compatible with the observer sampling varying Doppler factors when the outflow becomes optically thin. The multiple pulses then might be due to a similar origin from distinct persisting outflows, e.g., the observer samples a hollow conal geometry of a narrow structured jet as it sweeps past. Alternatively, these multi-pulses may be due to actual variability of energy injections from the magnetar crust and magnetosphere into the outflow. These timescales are compatible with characteristic oscillation mode periods of the magnetar, ranging from $\sim50-100$ ms for the lowest-order crustal torsional modes to milliseconds for $f$-modes. The latter is potentially important for GW searches of MGFs \citep{Macquet:2021eyn}. 

\subsection{Repeating magnetar probability calculation}
\label{repeating_prob}

GRB\,180128A occurred only 808 days before GRB\,200415A. The median volumetric rate of $3.8\times10^{5}$\,Gpc$^{-3}$\,yr$^{-1}$ \citep{burns2021identification} corresponds to 0.05 MGFs per 1.0\,M$_\odot$\,yr$^{-1}$. Given the star formation of NGC\,253  \citep{2019ApJS..244...24L} a predicted rate of 0.24 MGFs\,yr$^{-1}$ can be assumed. Excluding the assumption of uncorrelated events, the Poisson probability of producing these two MGFs in NGC\,253 in this time frame is $\sim$9.9\%, at 90\% confidence. Rejection of this assumption would suggest that a single, young magnetar is the source of MGFs well above the age-averaged MGF production rate for individual magnetars.

Nominally, this chance probability could be refined by accounting for the exposure and sensitivity in the observing instruments and the localization of GRB\,200415A containing only a portion of NGC\,253. Unfortunately, order of magnitude uncertainties currently dominate the intrinsic rate of MGFs. An unambiguous answer may be possible with more precise localizations of short GRBs. Below, we will detail the implications assuming the two events originate from a single magnetar. 

Lower energy magnetar short bursts are highly correlated in time (i.e., non-Poisson). No observations exist of two MGFs from the same magnetar in our Galaxy. A time interval of 808 days is short, yet does not definitively indicate that the two MGFs emanate from the same magnetar, as $\sim$2,190 days separated the SGR\,1900 and SGR\,1806 MGFs that came from different neutron stars. The magnetar crust is thought to be in a self-organized critical state \citep[e.g.,][]{1999ApJ...526L..93G,2022arXiv220908598L}; short bursts have a power-law event size distribution $d\log N/d\log E_{\rm iso} \sim -1.7$ \citep{1996Natur.382..518C} and have temporal correlations in activity rate similar to earthquakes and associated aftershocks \citep{2002PhRvL..88q8501B}. It is unclear if the trigger for MGFs is entirely different from that for short bursts or if they are identical and MGFs are merely much rarer and larger events. A tentative calculation in \cite{burns2021identification} suggests the short burst and MGF cumulative energy power-laws are compatible and could smoothly connect over $10$ orders of magnitude (within large uncertainties), favoring the scenario where the short burst and MGF physical triggering mechanism are fundamentally the same. This explanation could favor a more mature magnetar where the crust has had time to solidify and form strong stresses a few decades after formation \citep{2022arXiv220908598L} from presumably a SN.

If it is, in fact, the same magnetar, it would inform our understanding of the mechanism of MGFs and how stresses build and relax in the magnetar crust. Generally, magnetars in our Galaxy that have produced MGFs relax to less active states in the months or years following the event. The giant flares are energetic enough to putatively melt large zones of the outer crust, relieving stresses and large magnetic field twists. A second MGF would suggest a deep inner crust origin of the subsequent MGF trigger, matching expectations from some models \citep[e.g.,][]{2015MNRAS.449.2047L,2022ApJ...938...91K}. Statistics afforded by larger MGF populations are required to make definitive conclusions. Future proof of this would greatly benefit from the improved determination of the intrinsic rate uncertainties, as would newer high-energy wide-field monitors with arcminute scale localizations.

\section{Conclusions}
\label{sec:conclude}

Using the distinct MGF characteristic of prompt millisecond emission, paired with IPN localization to nearby star-forming galaxies, we have developed a robust and reproducible method for searching archival data for extragalactic MGFs weaker than those previously identified. From our initial results, GRB\,180128A stood out as a strong candidate MGF, supporting expectations that weak MGFs remain unidentified in archival data. Continued searches of archival \fermi\,data and the data of other missions may reveal yet more hidden MGFs. GRB\,180128A was subsequently localized to NGC\,253, making it the fifth likely MGF candidate localized to a nearby galaxy and the second such event localized to this specific galaxy. This localization marks the first time multiple MGFs have been found in a galaxy outside our own, supporting future studies to understand whether individual magnetars produce multiple MGFs. We have shown that the multi-pulse variability displayed in the lightcurves of GRB\,180128A, and other MGFs is a common trait. Our analysis of \textit{Fermi}-LAT data for GRB\,180128A did not detect the presence of \gev photons, contrary to the case of the MGF GRB\,200415A, directly showing that not all MGFs produce \gev photons.

Through our analysis and comparison of GRB\,180128A and GRB\,200415A, we have advanced our understanding of the physical mechanisms of these events. We extended the relativistic wind model put forth in \cite{2021Natur.589..207R} for GRB\,200415A, and by using selective binning techniques, resolve the finer structures within GRB\,180128A by changing the observational sampling of a MGF wind. Continued population studies will help to determine whether this model can reveal variation within the source class and whether such relativistic wind scenarios for MGFs are universal.

\section*{Acknowledgments}
\begin{acknowledgments}

AT, EB, MN, and OJR acknowledge NASA support under award 80NSSC21K2038. MN and ZW acknowledge the support by NASA under award number 80GSFC21M0002.
MGB thanks NASA for generous support under awards 80NSSC22K0777 and 80NSSC22K1576.
OJR gratefully acknowledges NASA funding through contract 80MSFC17M0022.
The \textit{Fermi}-LAT Collaboration acknowledges generous ongoing support
from several agencies and institutes that have supported both the
development and operation of the LAT and scientific data analysis.
These include the National Aeronautics and Space Administration and the
Department of Energy in the United States, the Commissariat \`a l'Energie Atomique and the Centre National de la Recherche Scientifique / Institut National de Physique Nucl\'eaire et de Physique des Particules in France, the Agenzia Spaziale Italiana and the Istituto Nazionale di Fisica Nucleare in Italy, the Ministry of Education, Culture, Sports, Science and Technology (MEXT), High Energy Accelerator Research
Organization (KEK), and Japan Aerospace Exploration Agency (JAXA) in Japan, the K.~A.~Wallenberg Foundation, the Swedish Research Council, and the Swedish National Space Board in Sweden. MASTER equipment was supported by Lomonosov MSU development program.\\

\end{acknowledgments}

\bibliography{GRB_bib}{}
\bibliographystyle{aasjournal}

\end{document}